\documentclass{andp2012}
\usepackage[english]{babel}
\title{Dynamical Many-body Localization and Delocalization in Periodically Driven Closed Quantum Systems}
\author[A. Haldar]{Asmi Haldar\inst{1,}\footnote{Corresponding author\quad E-mail:~\textsf{asmi.haldar@gmail.com}}}
\author[A. Das]{Arnab Das\inst{1}}
\address[1]{Indian Association for the Cultivation of Science\\
Department of Theoretical Physics \\
2A \& 2B Raja S. C. Mullick Road, Kolkata - 700032, India}
\begin{abstract}
Quantum interference lies at the heart of several surprising equilibrium and 
non-equilibrium phenomena in many-body Physics. Here we discuss two recently 
explored non-equilibrium scenarios where external periodic drive applied to 
closed (i.e., not attached to any external bath) quantum many-body systems have apparently opposite 
effects in respective cases. In one case it freezes/localizes a disorder 
free system dynamically, while in the other it delocalizes a disordered 
many-body localized system, and quantum interference is responsible for both the effects. 
We review these in the perspective of more general questions of ergodicity, energy 
absorption, asymptotic behavior, and finally the essential role of quantum mechanics 
in understanding these issues in periodically driven closed many-body systems. 
In this article we intend to deliver a non-technical account of some recent developments in this field
in a manner accessible to a broad readership. 
\end{abstract}
\shortabstract
\begin{document}
\maketitle

\section{Introduction}

Non-equilibrium dynamics of periodically driven closed quantum systems has gained significant 
attention recently, both because of its fundamental importance as a potential host of new quantum  
phenomena, and its recent experimental realizations (see, e.g., ~\cite{Bloch_Monika,Eckardt_Shengstock,
Mahesh_DMF_2014,Bloch_Ulrich}).
Few central issues, like energy absorption, thermalization, 
characterization of the asymptotic behavior of the system and role of quantum mechanics in
qualitatively understanding of these issues is overarching theme of this review. 
More specifically, we focus on two different settings where addressing these issues have lead to interesting 
phenomena and novel physical scenarios. Interest in periodically driven systems also 
stems from the possibility of generating topologically non-trivial phases (see, 
e.g.,~\cite{Polku_Bukov_Rev} for a review) but here we restrain ourselves from discussing those 
very interesting aspects.  

First, we review the phenomenon of dynamical many-body freezing (DMF) ~\cite{AD_PRB_2010} 
in presence of disorder~\cite{AR_AD_PRB_2015}. DMF is observed in a large class of translationally invariant 
integrable systems under strong and rapid periodic drive: the drive induces destructive quantum interference 
in a massive scale (i.e.,  affecting almost all degrees of freedom), and observables remain frozen 
close to their initial values for all time and for any arbitrary initial state. 
The picture here is, translational invariance and integrability allows one to map the 
many-body dynamics of these models to the population dynamics of a set of independent 
two-level systems, where it is possible to tune the drive parameters (frequency and amplitude) 
in such a way that strong destructive interference simultaneously affects all the modes. 
Introduction of disorder 
breaks translational invariance, rendering the above mentioned fine-tuning impossible, 
and observables eventually decays with time. Here it is worth noting that disorder is usually 
associated with localization and consequent freezing of dynamics, hence one needs 
to choose suitable observables and initial states in order to see the dynamics 
induced by disorder. Interestingly, however, even in presence of strong 
disorder, dramatic reminiscence of DMF still manifests itself:
an enormous enhancement of decay-timescale 
(orders of magnitude longer compared to the undriven case) is observed 
under the drive conditions corresponding to maximal freezing in the disorder-free 
systems. Thus in this case, periodic drive leads to freezing/localization in uniform 
systems and the disorder leads to unfreezing.   

Second, we review the effect of time-periodic drive on many-body localized 
systems~\cite{AL_AD_RM_PRL_2015,Jorge_MBL}. 
Here, disorder and interaction localizes a many-body system, and 
a periodic drive (unlike in the case of dynamical localization), delocalizes
the system. Though the phenomenology sounds more intuitive than that of DMF, 
its mechanism is subtly quantum mechanical.  
\begin{figure*}[ht]
	\begin{center}
  \includegraphics[width=0.4\textwidth]{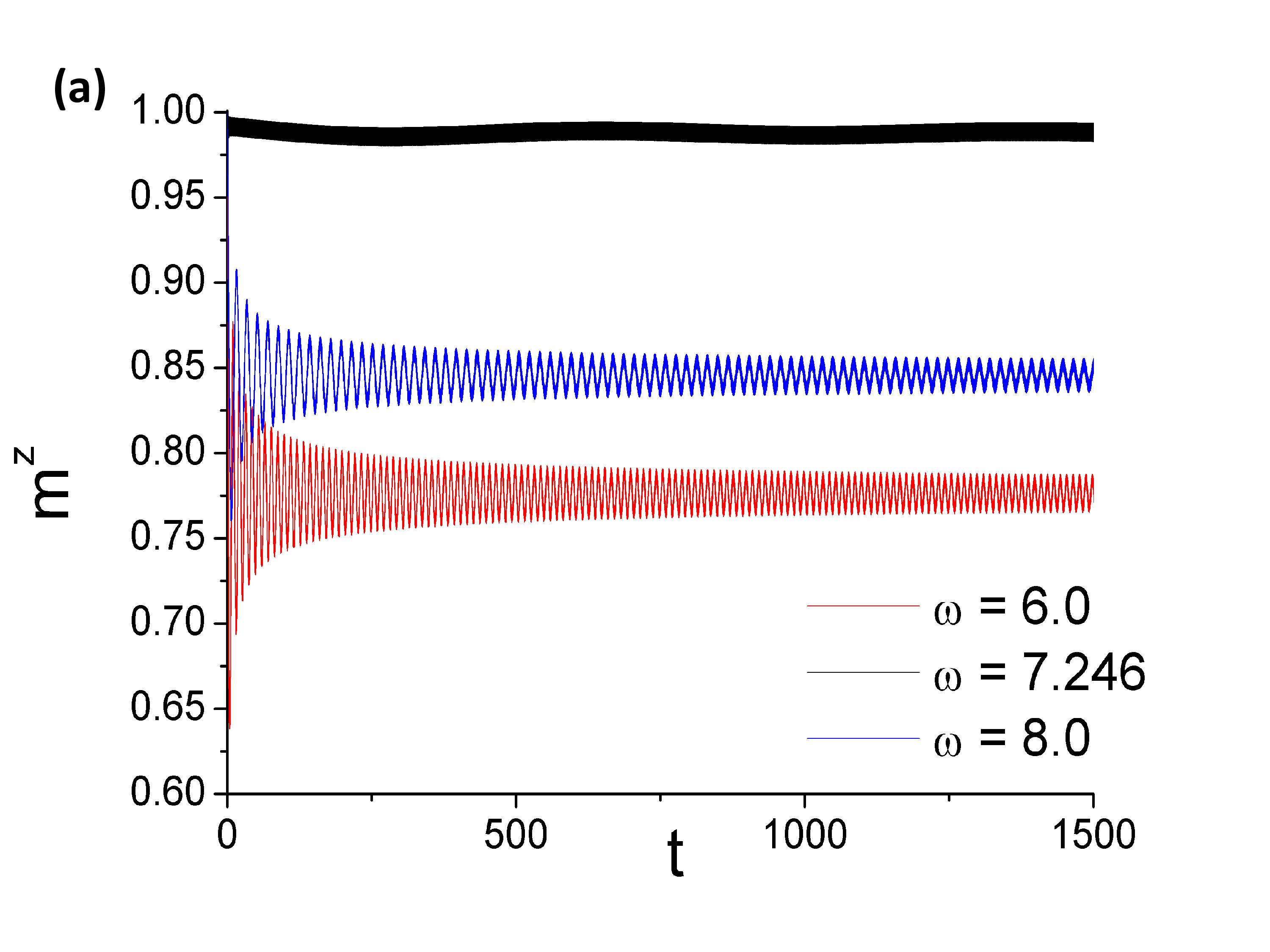}%
  \includegraphics[width=0.432\textwidth]{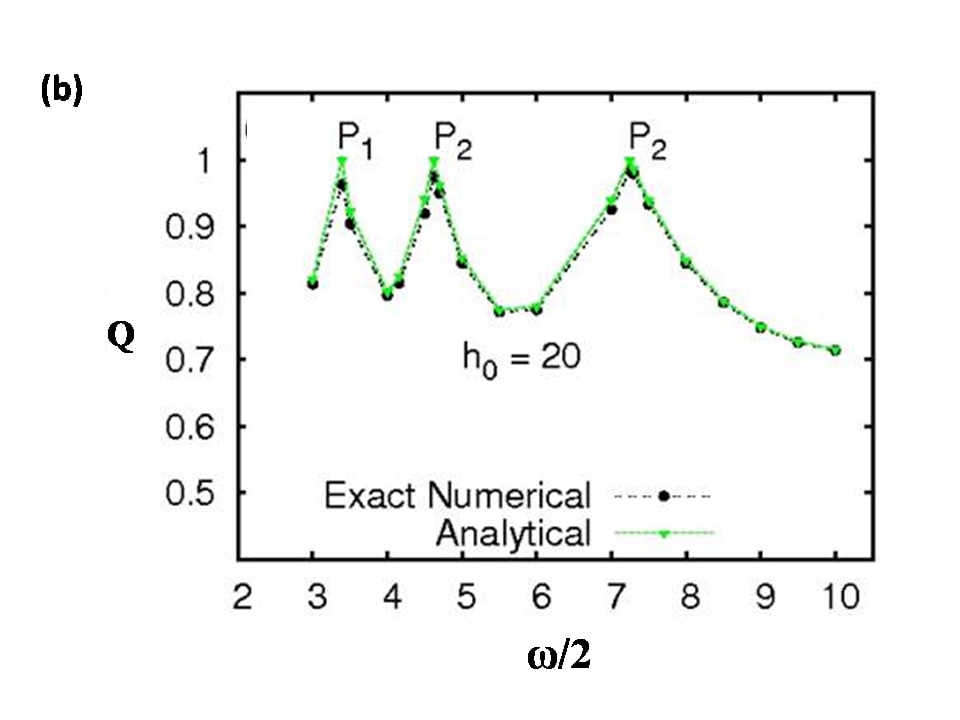}%
  \caption{Dynamical many-body freezing of transverse magnetization in 
	  homogeneous Ising chain in periodically 
	  driven transverse field. (a) $m^z$ vs $t$ (synchronization has not yet been attained within the t
	  timescale shown). (b) $Q$ (long-time average of $m^z$) vs $\omega$ - numerical and analytical results
	  under rotating wave approximation are compared. 
  (Fig. 1(b) taken from~\cite{AD_PRB_2010})} 
 \label{Fig_AD_PRB}   
	\end{center}
\end{figure*}
\section{Fate of Dynamical Many-body Freezing/Localization in Presence of Disorder}
It is known since long that periodically driven quantum systems with a single degree 
of freedom can undergo dynamical localization (or freezing) due to strong destructive 
quantum interference under certain drive conditions~\cite{Chirikov_Paper,Casati,Prange_PRL,Dunlap_Kenkre,CDT_PRL,CDT_Kayanuma_I,CDT_Kayanuma_II}. 
In some cases this happens even though the undriven system is quantum 
chaotic (i.e., the quantum system is obtained via quantization of a classically chaotic 
system). However, the physical picture that represents best the role of quantum 
interference in such localization phenomena can be quite diverse in different settings. 
For example, in case of a quantum kick-rotator, the mechanism of dynamical localization 
was identified~\cite{Prange_PRL,Prange_PRA} with that of Anderson 
localization~\cite{Anderson}, while in 
the case of a single particle moving on a plane~\cite{Dunlap_Kenkre} or in a 
double-well potential (coherent destruction of tunneling)~\cite{CDT_PRL,CDT_Kayanuma_I}, 
the mechanism of freezing seems to be viewed best as 
an effective dressing of the couplings/mass term in the Hamiltonian by the drive. An 
interesting connection between dynamical localization and coherent destruction of tunneling
has been revealed in ~\cite{CDT_Kayanuma_II}, and what happens to the phenomena in presence 
of interactions is an interesting open question.

A generalization of dynamical localization, namely dynamical many-body freezing (DMF) has 
been observed relatively recently in integrable translationally invariant quantum many-body 
systems~\cite{AD_PRB_2010,SB_AD_SDG_2012,AD_RM_Switching,Russomanno,Kris_DMF_2012,Polku_Bukov_Rev,Sei_Book} and has been realized experimentally~\cite{Mahesh_DMF_2014}. DMF is a 
generalization of the conventional dynamical localization 
in the following senses. First, in DMF the effect of {\it more than one 
(even mutually non-commuting) terms} in the  Hamiltonian can be simultaneously muted down via 
destructive interference induced by the drive. This is unlike the 
conventional dynamical localization where only one term (e.g., the kinetic energy) 
is suppressed. In simple cases, this might render strong freezing of certain 
observables for {\it any arbitrary initial state}, as 
the effective Hamiltonian responsible for the dynamics vanishes 
entirely~\cite{AD_PRB_2010,SB_AD_SDG_2012}. Second, DMF is a generalization 
of the conventional dynamical localization (observed for a single degree of freedom) to 
(infinitely) many-body systems (see, however, ~\cite{Andre_Eckardt_DL,Kris_DMF_2012,Sthitadhi_Amit_Diptiman_DL, Diptiman_Adhip_DL_1,Diptiman_Adhip_DL_2} for conventional dynamical 
localization in many-body systems studied more recently). It is to be noted that DMF 
is observed so far only in systems of non-interacting particles, or in those which can 
be mapped to one such. However, it is ``many-body" in the sense that it
survives even in the case when the dynamics is not factorizable to single-particle
sectors, i.e., it does not conserve the particle number. Moreover, DMF can be observed 
in presence of superconducting-like pair-creation/annihilation processes, which induces 
correlation between the particles (though not via a non-integrable interactions). 
Such inter-particle correlations are sufficient to drive long-range ordering and 
quantum phase transitions in many of these systems. Here we review the fate of DMF 
in presence of disorder~\cite{AR_AD_PRB_2015}. 

In ~\cite{AR_AD_PRB_2015}, the following disordered one-dimensional Ising chain 
subjected to a sinusoidal transverse field has been considered. The Hamiltonian is  
\begin{equation}
H(t) = -\alpha J \sum_{i}^{L} J_i \sigma^{x}_{i}\sigma^{x}_{i+1}  
-\sum_{i}^{L} \left\{h_{0}\sin{(\omega t)} + \alpha h_{i}\right\} \sigma^{z}_{i}, 
\label{H_Disord_AR_AD}
\end{equation}
\noindent where $\sigma^{\alpha}$'s ($\alpha = x,y,z$) are components of Pauli spins, $J_{i}$'s and 
$h_{i}$ are respectively the (quenched) interactions and on-site fields - both drawn randomly 
from a uniform distribution between $\left(-1,+1\right).$
The transverse field is subjected to an external  drive of frequency $\omega$ (period $T = 2\pi/\omega$) 
and amplitude $h_{0}$ ($\hbar = 1$, and periodic boundary condition). The study focuses on the regime of
{\bf strong} ($h_{0} \gg \alpha J J_{i}, \alpha h_{i}$) and {\bf fast} ($\omega \gg \alpha J J_{i}, \alpha h_{i}$) 
drive. One starts with the ground state at the initial Hamiltonian $H(t=0),$ 
drives the transverse field sinusoidally, and measures the transverse magnetization 
$m^{z}(t)$ as the response. \\

\noindent
{\bf DMF in the Homogeneous Chain (Fig.~\ref{Fig_AD_PRB}:} First we recapitulate the phenomenon of DMF
in absence of disorder 
($\alpha J J_{i} = 1$ and $h_{i}=0$ $\forall i$) ~\cite{AD_PRB_2010}. In this case $m^{z}(t)$ 
settles to a $T-$periodic state oscillating around a non-zero average value 
$Q = \lim_{\tau\to\infty} \frac{1}{\tau}\int_{0}^{\tau}m^{z}(t)dt$ for ever. Clearly, the system does not
absorb enough energy to destroy the order in the initial state however long one might drive 
(Fig.~\ref{Fig_AD_PRB} a). For dynamics with a fully polarized initial state, i.e., $m^{z}(0) = 1$, the magnitude  
of $Q$ can be used as a measure of freezing ($Q=0$ corresponds 
to the adiabatic drive and $Q = 1$ corresponds to infinitely rapid drive). As shown in
Fig~\ref{Fig_AD_PRB}(b), $Q$ turns out to be a highly non-monotonic 
function of the drive frequency $\omega$: $Q \approx 1/(1+{\cal J}_{0}(4h_{0}/\omega))$ 
(here ${\cal J}_{0}$ is the Bessel function of first kind of order 0, 
``$\approx$" denotes the rotating wave approximation). Thus under the condition 
\begin{equation}
	{\cal J}_{0}(4h_{0}/\omega) = 0,
\label{Frzn_Cond}
\end{equation} 
\noindent
the system freezes maximally, and one gets 
 $Q \approx 1$ (freezing peaks; labeled as $P_1, P_2$ etc in 
 Fig.~\ref{Fig_AD_PRB} b (we will refer to the condition in Eq.~(\ref{Frzn_Cond}) as 
 freezing peaks in the rest of the review).
\\ 

\noindent
{\bf Emergence of Stroboscopic Conserved Quantities and Role of Quantum Interference:} 
Existence of DMF clearly indicates absence of ergodicity and breakdown of Fermi-Golden Rule
type scenario even within the Hilbert space allowed by the inherent integrable structure of the model.
To be precise, integrability leads to decoupling of the degrees of freedom into independent two-level 
systems in momentum space. If the dynamics was ergodic for these two-level systems, each would keep 
on absorbing energy until it reaches an infinite temperature like state in the sense of having equal 
occupation probability for both energy levels, and one would have $\lim_{t\to\infty} m^{z}(t) = 0.$ 
The non-zero value of $Q$ or freezing is a consequence of repeated coherent interference of the amplitudes 
of the fundamental fermionic excitations in momentum space. 
If the interference effect is not taken into account, and the drive is assumed to change the population 
only according to the transition {\it probabilities}
(neglecting the interference between the transition {\it amplitudes}) after each cycle, the system
asymptotically approaches $m^z = 0$ state exponentially rapidly, rendering $Q=0$ regardless of the 
drive condition and initial state (illustrated in Suppl. Mat. of ~\cite{Mahesh_DMF_2014}).

The lack of thermalization (see, e.g.,~\cite{AD_PRB_2010,SB_AD_SDG_2012,AD_RM_Switching,
Russomanno}) can also be attributed to emergence of stroboscopic 
conserved quantities which leads to a periodic generalized 
Gibbs' ensemble (PGE) description of the system at long times~\cite{AL_AD_RM_PRL_2014}. 
The ensemble description of course involves concept of maximization of 
Von Neumann entropy in the Floquet space, which is equivalent to making the drastic assumption of equal apriori
probability taking into account only the relevant (stroboscopic) 
conservation laws. 

The above two standpoints of viewing the drive-dependent asymptotic states 
indicate the role of quantum interference - it determines the statistical ensemble
to which the system approaches asymptotically by fixing the relevant stroboscopic 
conserved quantities.

Quantum interference of course continues to play the overarching role also in the case of disordered systems
(since the above argument continues to hold), but the simples physical pictures necessary for intuitive 
understanding of the scenarios under different circumstances are quite diverse as we will see in 
the rest of the review.\\

\begin{figure*}[h]
	\begin{center}
  \includegraphics[width=0.95\textwidth]{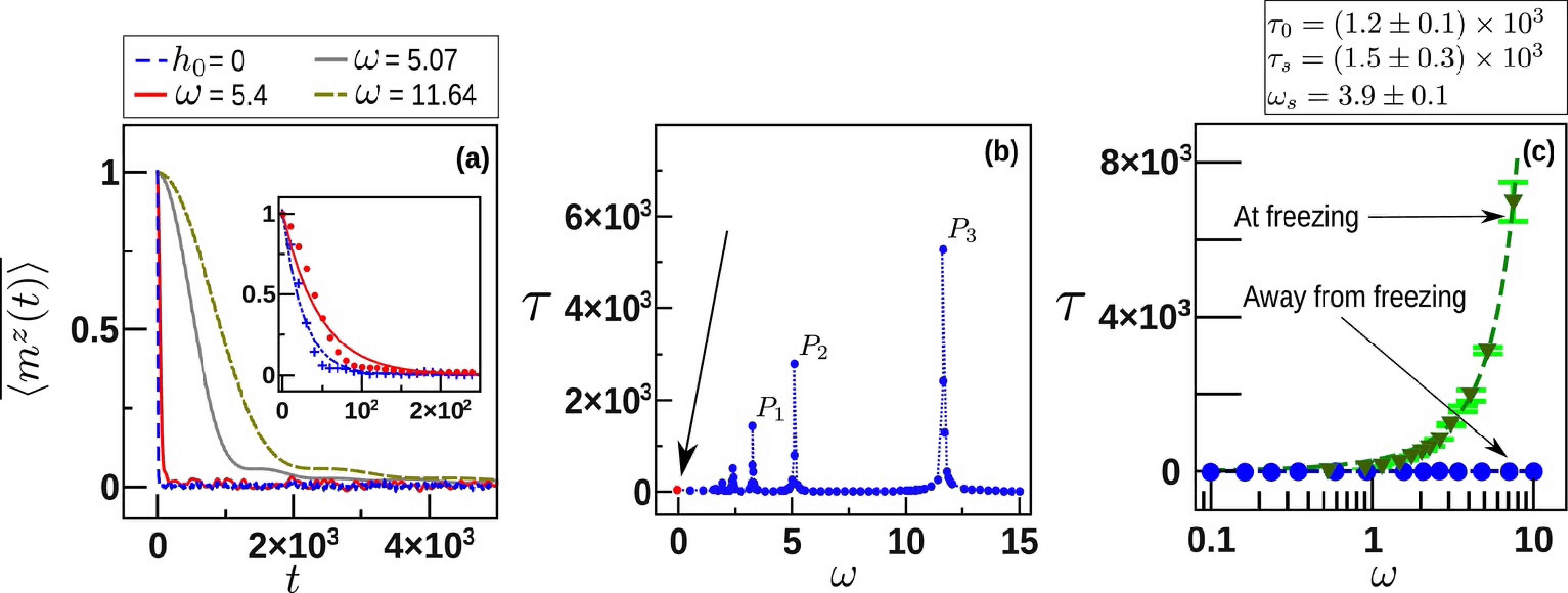}%
  \caption{Fate of dynamical localization in presence of random interactions without random fields.  
{\bf (a)} Exponential relaxation of the expectation value of ${m^{z}}$ with time for different
values of the drive frequencies $\omega$. Unless otherwise indicated, the drive amplitude $h_{0}$ 
is fixed at $7.0$. For certain specific values of $\omega$ (e.g., $\omega=5.07,11.64$), 
the relaxation is tremendously slowed down. The relaxation in the 
absence of the drive is labeled separately for comparison. The inset 
compares a representative sample of the numerical data (shown as points) 
to the curves fitted to them using Eq.~\ref{Exp-decay-fit}.  
{\bf(b)} {\it $\tau$ vs $\omega$ for fixed $h_{0}$}: 
The sharp peaks indicate dramatic enhancements of $\tau$ for certain values of $\omega$. 
Three of the most prominent peaks are identified
as $P_{1-3}$. The values of $\omega$ at these peaks are $P_1\approx3.24$, $P_2\approx5.07$, 
and $P_3\approx11.64$. Those 
values are identified to be the ones for which the effective Hamiltonian $H_{eff}$
vanishes. The red dot pointed by the arrow-head represents the 
case in absence of the drive. 
{\bf (c)} {\it $\tau$ vs $\omega$ at fixed $\eta$}:
Comparison of enhancement of $\tau$ as $\omega$ is increased keeping $\eta = \frac{4h_{0}}{\omega}$ fixed 
for two cases - under the freezing condition ${\cal J}_{0}(\eta) = 0$ ($\eta \approx 2.4048$), and away from it 
${\cal J}_{0}(\eta) \approx 0.765$
($\eta = 1.0$) as marked in the Fig. 
Exponential enhancement of $\tau$ with $\omega$ is observed (numerical data fitted with the
$\tau(\omega)\bigr|_{{J}_{0}(\eta)=0} =  \tau_{0} + \tau_{s}e^{{\omega}/{\omega_{s}}}$
form)  under the freezing condition, while no noticeable variation of $\tau$ is observed away from the
freezing condition. 
Results are for $L=100,$ averaged over $> 10^3$ 
disorder realizations of the bonds $J_i$.
The error-bars due to disorder-induced fluctuations are about the point size, hence omitted. 
	  Qualitatively similar results are observed with random fields. (Fig 
	  taken from ~\cite{AR_AD_PRB_2015})
  }
 \label{Fig_AR_AD_PRB_1}   
	\end{center}
\end{figure*}
\noindent
{\bf Unfreezing by Disorder and Strong Remnants of DMF: The Phenomenology} 
In presence of disorder, $m^z$ always 
decays to zero regardless of the drive parameters at infinite time and DMF is eventually
destroyed. The decay can be fitted well with the exponential decay form (Fig. ~\ref{Fig_AR_AD_PRB_1}(a))
\begin{equation}
\overline{\langle m^{z}(t)\rangle} =  m^{z}_{0}e^{-t/\tau},
\label{Exp-decay-fit}
\end{equation} 
\noindent 
where the overbar denotes average over disorder realizations. However, the decay time-scale $\tau$ depends 
dramatically on the drive parameters $(h_{0},\omega),$ and exhibits a spectacular reminiscence 
of DMF. As shown in Fig. ~\ref{Fig_AR_AD_PRB_1}(b), the relaxation 
time scale $\tau$ shoots up by {\it several orders of magnitude} when the freezing condition for
the uniform chain (Eq.~\ref{Frzn_Cond}) is satisfied by the drive. Interestingly, if $\eta$ is kept
fixed to a value such that ${\cal J}_{0}(\eta) = 0,$ (corresponding to peak freezing in the homogeneous 
system) and $\omega$ is increased, then $\tau$ 
increases exponentially with it (see Fig.~\ref{Fig_AR_AD_PRB_1}(c)). 
Thus, though introduction of disorder eventually kills the freezing of 
$m^{z},$ the timescale of decay still bears a very strong signature of the extreme freezing 
observed in the absence of disorder. \\ 

\noindent
{\bf The Points of Maximal Freezing from Floquet Flow Equation Approach:} It seems difficult to find 
analytical solution of time-dependent Schr\"{o}dinger equation with disordered Hamiltonians. 
Hence one can resort to the following Floquet analysis and determine the effective Floquet Hamiltonian 
approximately using a flow equation approach~\cite{Mintert}.
For the present purpose, we adopt the simplest formulation of Floquet theory and define the effective
time-independent Hamiltonian that describes the evolution of the stroboscopically observed wave-function as follows.
Let us denote the time evolution operator evolving a state through a period from
$t=\epsilon$ to $t=\epsilon + T$ ($0  \le \epsilon < T$) by $U (\epsilon)$. 
Since $U(\epsilon)$ is unitary, it can always be
expressed in terms of a hermitian operator $H_{eff}$ as
\begin{equation}
U(\epsilon) = e^{-iH_{eff}(\epsilon)T}. 
\label{Heff_def}
\end{equation}
\noindent
Clearly, if observed in a ``stroboscopic" fashion at instants 
$t = \epsilon, \epsilon + T, \dots, \epsilon + nT$ ($n$ is an integer), the dynamics 
can be considered to be effectively governed as if by a time-independent Hamiltonian 
$H_{eff}.$ With $H_{eff}$ one gets the same wave-function as that with $H(t)$ at
the instants $t = \epsilon + nT,$ because the time-evolution operator is same 
in both cases for evolution to those instants 
($[e^{-iH_{eff}T}]^{n} = e^{-iH_{eff}nT}$). This of course holds for every $\epsilon,$ 
hence we get different stroboscopic series for each of them (actually choice of $\epsilon$
is equivalent to choosing a gauge, as shown in~\cite{Polku_Bukov_Rev}).
For characterizing the long-time behaviour of the system under
rapid drive, it is sufficient to observe the system strobocopically, 
since nothing much happens within a single cycle. Hence it is sufficient to 
follow the dynamics governed by $H_{eff}(\epsilon=0)$ at $t = nT.$ Moreover,
the set of all (i.e., for all values of $\epsilon$) stroboscopic 
observations are sufficient to construct the entire time evolution (see 
~\cite{Polku_Bukov_Rev} for an elegant and efficient way of extracting this information).

The Hamiltonian in Eq.~\ref{H_Disord_AR_AD} can be mapped to the following non-interacting
Hamiltonian using standard prescription~(see, e.g., \cite{Sei_Book}).
\begin{multline}
H(t) = - \alpha J \sum_{i}^{L}  J_i \left(c^{\dagger}_i c^{\dagger}_{i+1} 
+ c^{\dagger}_i c_{i+1}  + {\rm h.c.}\right) \\
- 2 \sum_i^{L} \left\{h(t)+\alpha h_i\right\} c^{\dagger}_i c_i,
\label{H_Fermi}
\end{multline}
\noindent
with hard-core bosons  created (annihilated) by $c^\dagger_i$ ($c^{\;}_i$)
These bosons satisfy $\{ c_{j}^{\dagger},c_{j} \}=0$, and the usual bosonic commutation relations for $i\ne j$. 
Also, $h(t)=h_0\sin{\omega t}$. Using Floquet-flow equation technique, one can construct the 
following analytical expression for $H_{eff}$ for the evolution from $t=0$ to $t=T$ governed by the Hamiltonian in 
Eq.~\ref{H_Fermi}~\cite{AR_AD_PRB_2015}.
\begin{multline}
\label{eq:heff}
H_{\rm eff}    \approx   -J\sum_i j^{(0)}_i\left(c^\dagger_i c^\dagger_{i+1} + \rm{h.c.}\right) \\
		            -J\sum_i j^{(1)}_i\left(c^\dagger_i c^{\;}_{i+1}+\rm{h.c}\right) 
		            -\mu\sum_i j^{(2)}_i c^\dagger_i c^{\;}_i,
\end{multline}
with $\eta \equiv 4h_0/\omega$, and the constants $j^{(s)}_i, \mu $ defined as follows.
\begin{eqnarray}
	j^{(0)}_i            &\equiv& \alpha J_i\; \bigg\{{\cal J}_0(\eta) - \frac{4\alpha h_i}{\omega}\beta(\eta)\bigg\} ,\nonumber \\
	j^{(1)}_i            &\equiv& \alpha J_i \; {\cal J}_0(\eta) ,\nonumber \\
j^{(2)}_i            &\equiv& \frac{h_i}{J},\nonumber \\
\mu 		     &\equiv& 2\alpha J.
\label{Renorm_fctr}
\end{eqnarray}
Here, ${\cal J}_n(\eta)$ denote Bessel function of the first kind of order $n,$
and $\beta(\eta)\equiv\sum_{n\neq 0} {{\cal J}_n(\eta)}/{n}$. 
The above is obtained under a rotating-wave approximation (RWA) which holds for
$\omega \gg J,\alpha.$
This effective Hamiltonian accurately reproduces the dynamics 
of the full system stroboscopically to the leading order in 
$\alpha/\omega.$ 

The above expression of $H_{eff}$ shows that when ${\cal J}_{0}(\eta) = 0$ (freezing peaks) 
and $h_{i} = 0,$ $H_{eff}$ vanishes, implying complete freezing of the dynamics. 
Though this does not happen here and $m^{z}$ decays to zero due to the effect 
of the higher order terms in $\alpha/\omega$ (unlike in the case of the homogeneous 
chain), the time-scale $\tau$ gets enormous jumps at these points (see Fig.~\ref{Fig_AR_AD_PRB_1} a,b). 
For $h_{i} \ne 0$ the picture is qualitatively similar, though the enhancements 
of $\tau$ are smaller due to the presence of the $\beta$-term. Note that the huge 
enhancement of time-scale is achieved since the drive strongly suppresses three 
different mutually non-commuting terms in the Hamiltonian - a hallmark of DMF.

At the freezing peaks, $\tau$ can also be enhanced exponentially by tuning $\omega ,$ (Fig.~\ref{Fig_AR_AD_PRB_1}c),
demonstrating a great control achievable on the disordered induced decays via periodic drive.

Before concluding this section a few words on RWA seems to be in order. In RWA, one essentially goes into a
``rotating frame" where the sinusoidal drive term in the Hamiltonian can be expanded 
into sum of terms of the form $e^{(i\Omega t)},$ and drops out all the terms for which the 
frequency $\Omega$ is much larger than the characteristic frequencies of the undriven system.
DMF occurs when the drive frequency is high enough to be 
off-resonant with all the characteristic frequencies of the system (i.e., in the limit of zero-photon process). 
In the present case, this condition translates to $\omega \gg J,\alpha$~\cite{AR_AD_PRB_2015}. A detailed 
and critical review on the domain of validity of RWA is given in~\cite{Ashhab}.

\section{Fate of Many-body Localization under Periodic Drive}

In this part of the review we take a different standpoint: We consider interacting systems where disorder
induces many-body localization (MBL) in the Fock space (see, 
e.g.,~\cite{Anderson,Altshuler,Imbrie_PRL,Abanin_lbits,Huse}), 
and the question is if one can delocalize the system by applying an external periodic drive. But before
addressing this, we first make a small detour and briefly review the application of the Floquet 
formalism in studying the asymptotic properties a periodically driven many-body systems in general. \\

\subsection{Effective Floquet Hamiltonian and the Diagonal Ensemble}
\label{Diagonal_Ensemble}
Consider a static Hamiltonian $H_{0}$ hosting a many-body localized phase be driven by a
time-periodic (non-commuting with $H_{0}$) part $H_{D}(t).$ The total Hamiltonian is thus
\begin{equation}
H(t)=H_{0}+H_{D}(t),\label{eq:H0-hd}
\end{equation}
\noindent and
$H_{eff}\left(\epsilon\right)$ be the corresponding Floquet Hamiltonian (see Eq.~(\ref{Heff_def})). Then 
\begin{equation}
\exp\left(-iH_{eff}\left(\epsilon\right)T\right)
= \mathcal{T}\exp\left(-i\int_{\epsilon}^{\epsilon+T}dt\: H(t)\right),\label{eq:defn-heff}
\end{equation}
\noindent where ${\mathcal T}$ denotes time-ordering. 
Without loss of generality one can set $\epsilon=0.$
Let $|\mu_{i}\rangle$ denote the $i$-th eigenstate of 
$H_{eff}$ corresponding to the eigenvalue $\mu_{i}$. \\
Under ``generic" initial conditions and considering ``generic" local operators 
as observables, the nature of the asymptotic state can be understood from the following. 
In order to consider the fate of the system at long times, we consider a initial state
$$|\psi(0)\rangle = \sum_{i}c_{i}|\mu_{i}\rangle$$ 
\noindent
and an observable 
$$\hat{\cal{O}} 
= \sum_{i,j}{\cal O}_{ij}|\mu_{i} 
\rangle\langle \mu_j|.$$
\noindent
\begin{equation}
	\langle \psi(nT+\epsilon) |{\hat {\cal O}}|\psi(nT+\epsilon) \rangle 
	= \sum_{i,j}c_{i}c_{j}^{\ast}{\cal O}_{ij}e^{-i(\mu_{i}-\mu_{j})(nT+\epsilon)}. 
\label{O_exp}
\end{equation}
\noindent
For a many-body system (infinite-size limit), by ``generic" $|\psi(0)\rangle$ and by ``generic" operator 
${\hat{\cal O}},$ we mean that above sum is extensive in the sense that there are sufficiently 
large number of quasi-energy states participating in the sum. In that case (see ~\cite{Riemann,AL_AD_RM_PRL_2014} 
for further conditions), at long times ($n\to\infty$) the off-diagonal ($i\ne j$) terms in the sum oscillates 
rapidly and their contributions add up almost randomly, summing up to a vanishingly small quantity. 
Hence under above conditions, the state of the system 
can be described by an effective ``diagonal ensemble" given by the mixed density matrix~\cite{Rigol_Nature}
$$
{\hat \rho}_{_{Diag}} = \sum_{i}|c_i|^2 |\mu_{i}\rangle\langle\mu_{i}|.
$$
\noindent
Thus, the asymptotic properties of a periodically driven system are 
effectively given by a statistical average over the expectation values of the eigenstates of $H_{eff}$,
and hence it is sufficient to study the nature of the eigenstates and eigenvalues of $H_{eff}$ in order
to understand the long-time behaviour. Moreover, reduction of the stroboscopic dynamics to that due to
a time-independent $H_{eff}$ implies that quasi-energy plays similar role in the stroboscopic 
dynamics as energy plays in the dynamics governed by a time-independent Hamiltonian.
These hold regardless of whether the system is 
ergodic or many-body localized, see, e.g.,~\cite{Goold_Gogolin_Scardicchio_Silva}. 
Here it is worth noting that in spite of this reducibility, nature of the dynamics due to periodic drive
can be different from a generic undriven case in fundamental ways, since in the former 
case $H_{eff}$ might be highly non-local. For example, dynamics under a generic time-independent 
local Hamiltonian leads to thermalization at a finite temperature, while evolution under an $H_{eff}$ 
derived from a time-periodic generic Hamiltonian can lead to heating up to an effectively infinite temperature 
scenario~\cite{AL_AD_RM_PRE_2014,Alessio_Rigol_PRX}.

\subsection{Delocalization of MBL} 
MBL is a thermodynamically stable non-ergodic phase of matter (see, e.g., ~\cite{Anderson,Altshuler,Imbrie_PRL,
Abanin_lbits,Huse}) where an interacting many-body system remains localized in the Fock space due to disorder in
absence of an external bath. Here we address if periodic drive can destabilize such a phase and heat it up indefinitely. 
From the discussion in Sec~\ref{Diagonal_Ensemble} it is clear that in order to distinguish between an 
MBL and an ergodic phase in a periodically driven system, it is sufficient to focus on the properties/statistics of 
eigenstates and eigenvalues of $H_{eff}$~\cite{AL_AD_RM_PRL_2015}.
The general scenario depends on the absence (presence) of many-body mobility edge as summarized below. 
Two mechanisms by which periodic driving might destroy MBL are identified. 
The first, rather robust, mechanism is the mixing of undriven eigenstates 
from everywhere in the spectrum by the driving; if there is a mobility edge, this results in delocalization 
of all states of the effective Hamiltonian. The second mechanism is more subtle and involves strong mixing 
of states~\cite{AL_AD_RM_PRL_2014} which causes a delocalization transition at a {\it finite} drive frequency
for a given disorder strength.  
The key findings are summarised in Table~\ref{tab:Effect-of-driving}.
\begin{table}[h!]
	\begin{tabular}{|c|c|c|}
	\hline
	\textbf{Mobility edge} & low frequency & high frequency\tabularnewline
	\hline
	\hline
	present & delocalized & delocalized\tabularnewline
	\hline
	absent & delocalized & \textbf{localized}\tabularnewline
	\hline
	\end{tabular}

	\protect
	\caption{Effect of driving frequency in the presence
		and absence of a mobility edge \label{tab:Effect-of-driving}
	}

\end{table}

\begin{figure}[h]
	\begin{center}
  \includegraphics[width=0.4\textwidth]{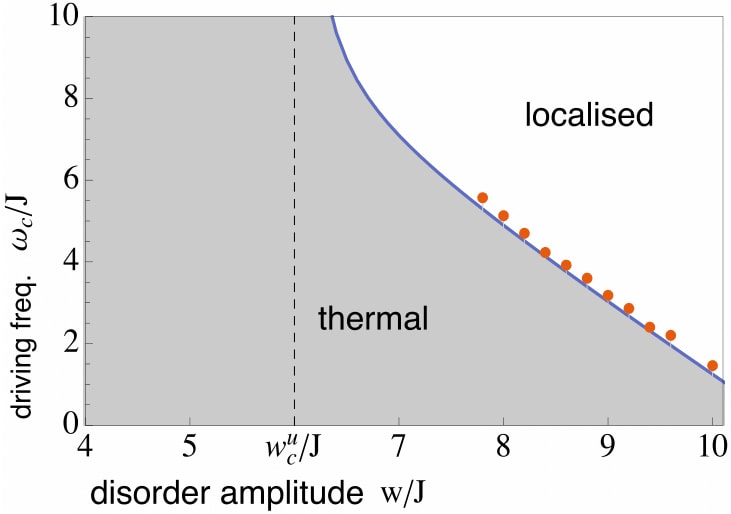}%
  \caption{Driven MBL with no mobility edge: Plot of driving frequency $\omega_{c}$ below which the system delocalizes
as a function of disorder amplitude $w$. The shaded areas correspond
to delocalization. The red dots are obtained from finite-size studies
of the level statistics of the system. The disorder amplitude $w_{c}$
is the value below which the undriven system is delocalized in the
absence of driving. The blue line is a guide to the eye. (Fig. taken from~\cite{AL_AD_RM_PRL_2015})
} 
 \label{fig:omegac-vs-w}   
	\end{center}
\end{figure}

This leads to a phase diagram outlined in Fig~\ref{fig:omegac-vs-w}. For low enough
$\omega$ and disorder strength, the system always delocalizes under the drive,
while for high enough $\omega$ and disorder the system remains MBL.
The blue line in Fig.~\ref{fig:omegac-vs-w} indicates a tentative boundary 
between these two phases, obtained by extrapolating the numerically determined transition points (red dots).
\noindent
In the following two different models (with/without the mobility edge) are considered separately in order to
illustrate the phenomenology above. \\

\subsubsection{No mobility Edge} 
\label{No_Mobility_Edge}
A model of interacting hard-core bosons is considered,
which is described by a driven, 
local Hamiltonian (Eq.~\ref{eq:H0-hd}) with
\begin{equation}
	H_{0} =
		H_{hop}
		+\sum_{r=1}^{2} 
			V_{r}\sum_{i=1}^{L-1}n_{i}n_{i+r}
		+\sum_{i=1}^{L}U_{i}n_{i}\label{eq:H}
\end{equation}
where $H_{hop}=\left(-\frac{1}{2}J\sum_{i=1}^{L-1}\left(b_{i}^{\dagger}b_{i+1} 
+ b_{i+1}^{\dagger}b_{i} \right)\right)$ is a hopping operator, the $b$ 
are hard-core bosonic operators, $U_{i}$ an on-site random potential uniformly 
distributed between $-w$ and $+w$ and $H_{D}\left(t\right)$ a time-periodic hopping term
\begin{equation}
	H_{D}\left(t\right)=\delta\tilde{\delta}(t) H_{hop}
	\label{eq:periodic-hopping}
\end{equation}
with $\delta$ a dimensionless constant, $\tilde{\delta}(t)=-1(+1)$ in the first (second) half of each period $T=2\pi/\omega$. 
Via Jordan-Wigner transformations this model is related to a fermionic interacting system as well as to a spin-1/2 chain. 
The results are presented for $V_{1}/J=V_{2}/J=1$, although the qualitative conclusions are not 
sensitive to this. \\ 

\begin{figure}[h]
	\includegraphics[scale=0.4]{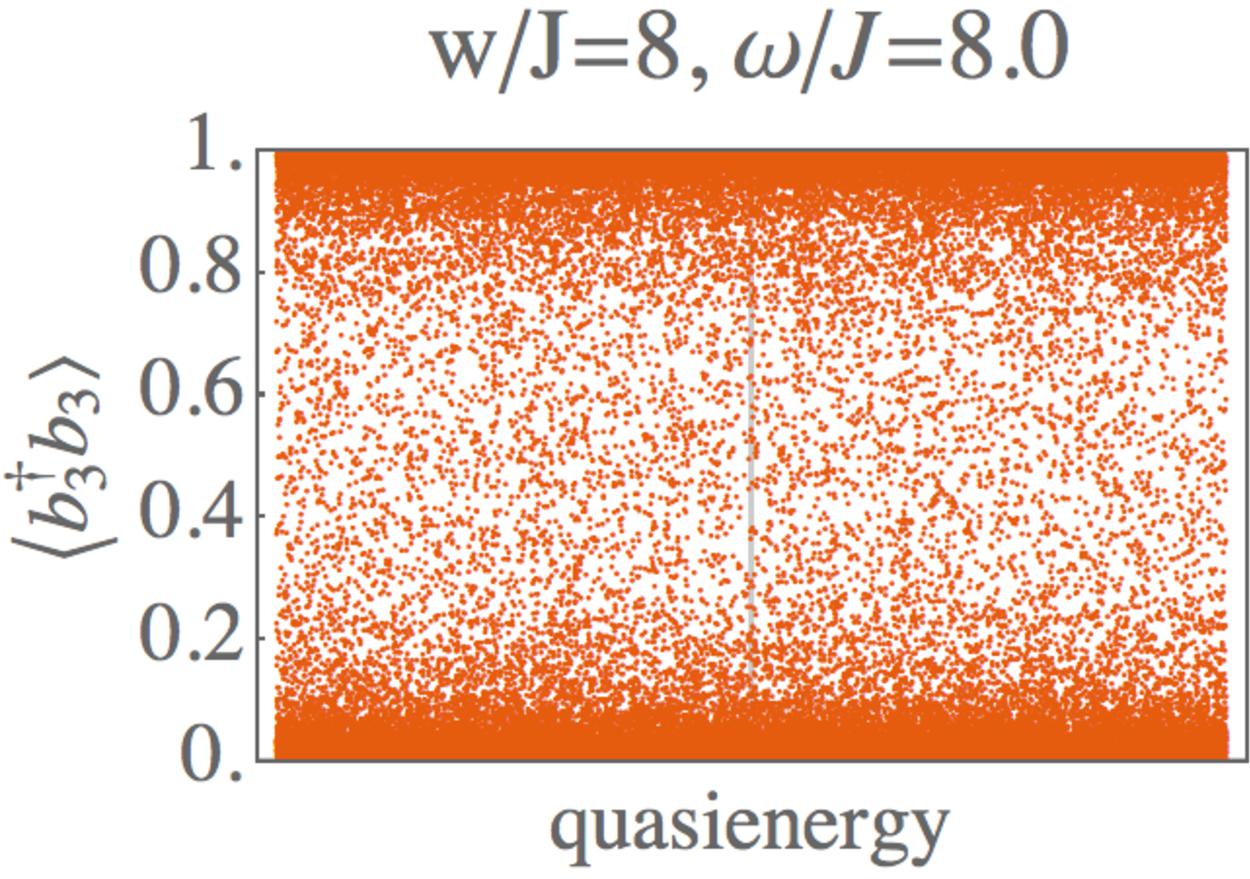}
	\includegraphics[scale=0.4]{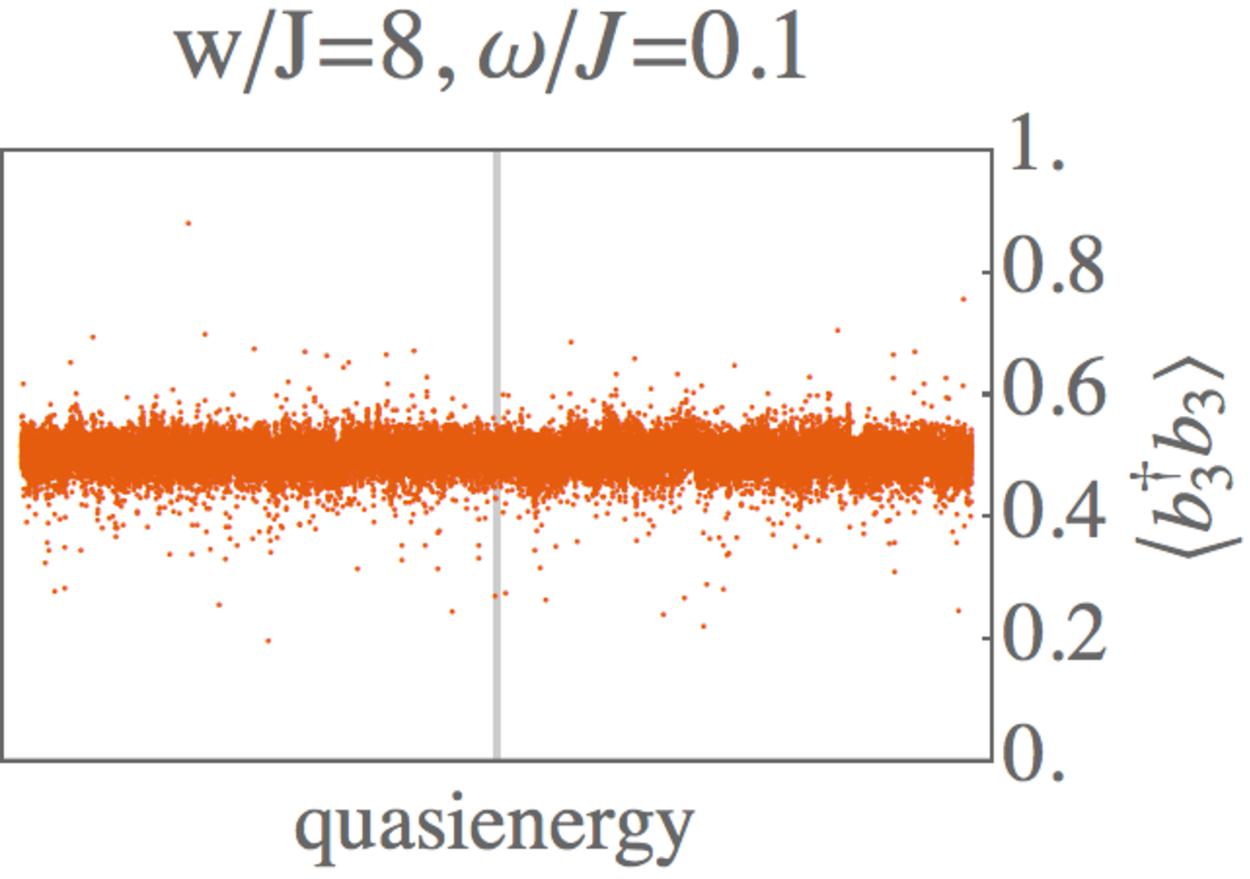}
\protect\caption{Plots of eigenstate expectation values (EEV) of the density at a single
arbitrarily chosen site in all the eigenstates of $H_{eff}$ for a
system with $w/J=8.0$, size $L=18$ for a Hilbert space dimension
of $D_{H}=48620$ and driving amplitude $\delta/J=0.1$. For driving
frequency above the blue line in Fig.~\ref{fig:omegac-vs-w}, $\omega/J=8.0$
(left), the EEVs fluctuate wildly between different eigenstates of
$H_{eff}$. In contrast, for a driving frequency below the blue line,
$\omega/J=0.1$ (right), there is markedly less eigenstate-to-eigenstate
variation, consistent with all states being fully mixed. This is the
expected behaviour of the EEVs for clean (therefore delocalized) driven
systems (see Ref.~\cite{AL_AD_RM_PRE_2014}). In the undriven system
the EEVs appear qualitatively similar to those in the left panel. 
(Fig. taken from~\cite{AL_AD_RM_PRL_2015})
\label{fig:Plots-of-EEVs}}
\end{figure}

\begin{figure*}
	\begin{center}
\includegraphics[scale=0.50]{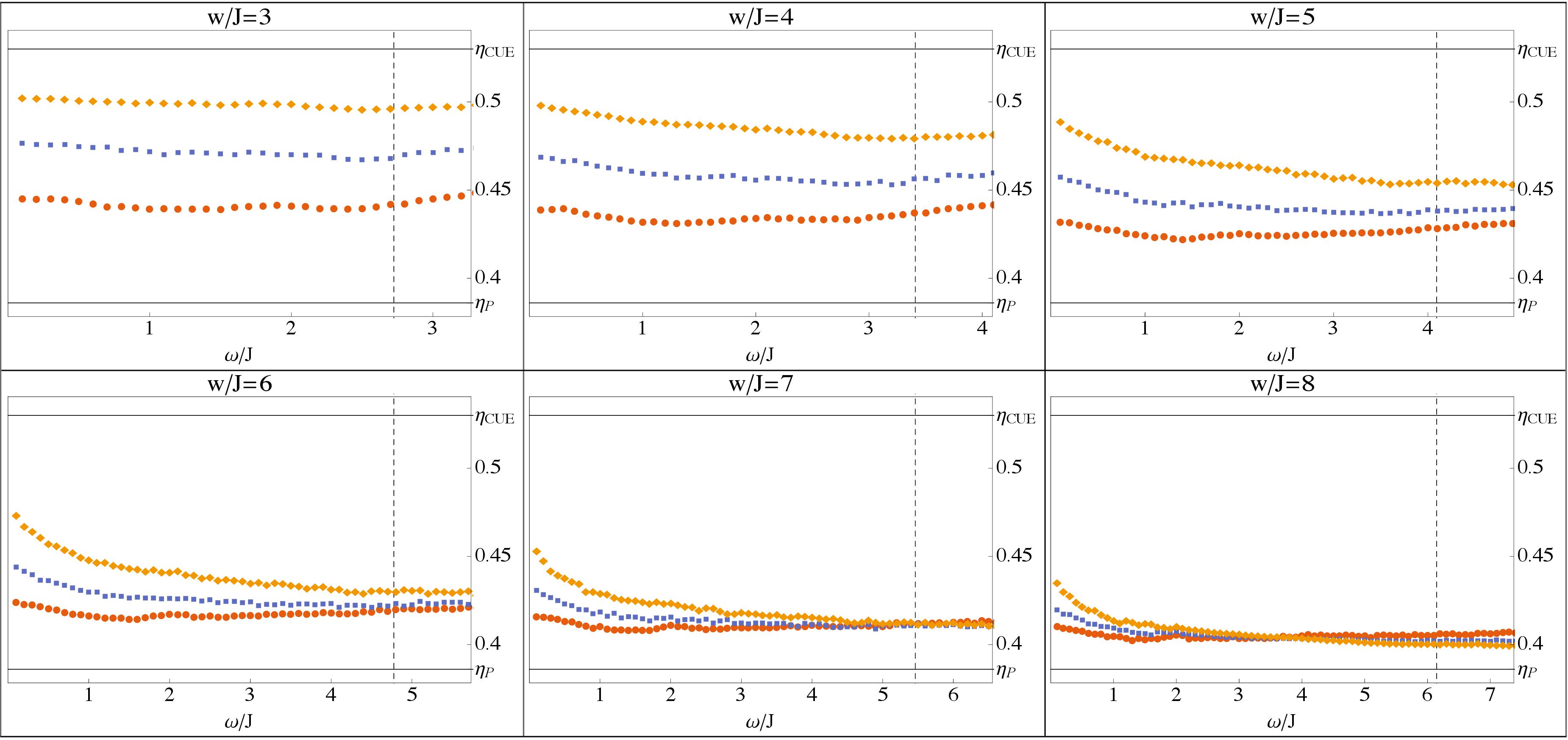}
\protect\caption{Level statistics for various disorder amplitudes $w/J$ as a function
of driving frequency $\omega$. The driving amplitude is $\delta/J=0.1\ll w/J,\omega/J$,
and each point represents an average over 10000 disorder realisations.
The dashed vertical lines indicate half the width of the energy spectrum;
for $\omega$ greater than this, the results cannot be extrapolated
to the thermodynamic limit. The colours correspond to system sizes 
$L=8,10,12$ from bottom to top for the smallest $\omega$. The values 
$\eta_{CUE}$ and $\eta_{P}$ correspond to the presence and absence of 
level repulsion respectively, which in turn correspond to localized 
and delocalized phases. The dotted vertical lines correspond to the 
typical spectral width of the system, for frequencies above which the results cannot be used
to infer the thermodynamic limit. (Fig. taken from~\cite{AL_AD_RM_PRL_2015}).
\label{fig:example-level-stats}}
	\end{center}
\end{figure*}

\noindent{\bf Non-ergodicity of the MBL Phase from Eigenstate Expectation Values (EEV):}
It has been shown that generic interacting systems, when driven periodically in time
at low frequencies (low compared to the bandwidth of the undriven system), the system
keeps on absorbing energy without bounds, ending up in a state which is indistinguishable
from an infinite temperature state as far the expectation values of local observables are
concerned~\cite{AL_AD_RM_PRE_2014,Alessio_Rigol_PRX}. This infinite temperature like 
scenario is reflected in the fact that the eigenstate expectation values (EEV)
of any local operator 
$\langle {\hat{\cal O}} \rangle_{i} = \langle\mu_{i}| {\hat{\cal O}} |\mu_{i}\rangle$ 
over all eigenstates of $H_{eff}$ are almost equal to each other (i.e., when EEV is plotted with 
respect to $\mu_{i}$ it is almost flat for any given ordering of $i$). On the other hand,
in the localized phase EEV fluctuates wildly.  
This is illustrated in Fig.~\ref{fig:Plots-of-EEVs}. \\

\noindent {\bf Locating the MBL-Delocalization Transition:} 
To accurately locate the localization-delocalization transition for
the driven system, the level statistics of the eigenvalues
of $H_{eff}$ has been calculated. That is,
after obtaining the quasi-energies $\epsilon_{n}$, one calculates
the following ratio involving adjacent level spacings 
$\delta_{n}=\epsilon_{n}-\epsilon_{n+1}$: 
$r_{n}=\min\left(\delta_{n},\delta_{n-1}\right)/\max\left(\delta_{n},\delta_{n-1}\right)$. 
The mean $\eta=\int_{0}^{1}dr\, rP(r)$
distinguishes between Circular Unitary Ensemble (ergodic) 
and Poissonian statistics (non-ergodic). One calculates $\eta$ for a sequence 
of system sizes and extrapolate the limit of $\eta$ as $L\rightarrow\infty,$ 
and consider the statistics of the extrapolated values of $\eta.$ 

To obtain the frequency $\omega_{c}$ above which delocalization 
sets in for a driven system, one plots $\eta$ for several values 
of disorder amplitude $w$, averaged over $\sim 10^4$ disorder 
realisations and for several system sizes. Typical results are shown
in Fig.~\ref{fig:example-level-stats}. The transition is located
at the crossing point of the lines for different system sizes: if
increasing system size results in larger $\eta$ then we conclude
that the system is delocalized, since $\eta=\eta_{CUE}$ for a delocalized
system and $\eta=\eta_{P}$ for a localized system with $\eta_{P}<\eta_{CUE}$. 
Here $\eta_{CUE}$ is the value for the CUE ensemble~\cite{Alessio_Rigol_PRX}.

For the results to be applicable in the thermodynamic limit,
it is necessary (though might not be sufficient) to take 
the drive frequency $\omega$ much lower than the width of the energy
spectrum of the undriven Hamiltonian. 
The main practical problem is the following: with decreasing
disorder amplitude $w$ and for fixed system size, the value $\omega_{c}$
increases while the energetic width of the DOS decreases.
Since $\omega$ must be small compared to the width in order for the
extrapolation to the thermodynamic limit to be meaningful, the $\omega_{c}$
for values of the disorder close to $w_{c}$ are inaccessible for
the system sizes for which the numerics could be done. The width of the DOS is indicated
in Fig.~\ref{fig:example-level-stats} by vertical lines; the crossing
point of the curves cannot lie to the right of this line, since otherwise
the finite size of the system would be important (and thus the results
would not be reliable in the thermodynamic limit).

Fig.~\ref{fig:example-level-stats} reveals the following
features: for $w/J\leq 6$ (where the undriven system is delocalized)
the lines for successive, increasing $L$ do not cross for values of
$\omega$ below the bandwidth, indicating that the thermodynamic limit
is delocalized, as expected. For $w/J>6$, there is a clear crossing
point, which indicates the position of the transition. The crossing
value of $\omega$ determined by this method is plotted as a function
of $w/J$ in Fig.~\ref{fig:omegac-vs-w}. \\

\noindent
{\bf A Physical Picture:} Though the physical picture consistently fitting the 
full phenomenology described above is not entirely clear, significant efforts 
have been made in this direction~\cite{AL_AD_RM_PRL_2015,Abanin_MBL,Abanin_Anushya_Annals}. 
We summarize the gist in the following. \\

\noindent
{\bf Delocalization Under Low-frequency Drive:} In the MBL phase and in the absence of driving, 
the system is effectively integrable in that there exist extensively many local integrals of 
motion~\cite{Abanin_lbits,2013PhRvL.111l7201S,1407.8480,Ros:2014vz,Imbrie:2014vo}. 
The system may thus be thought of as a set of \emph{local} subsystems, of finite spatial extent.
The matrix elements connecting these local systems are suppressed in much the same way hopping
amplitude between sites is suppressed in Anderson localization. Under slow periodic drive, the
MBL Hamiltonian is replaced by an $H_{eff},$ which is not just the time-average of the periodic 
Hamiltonian over a period, but also have significant other components which consists of 
effective long-range hopping and interactions. These terms might introduce matrix elements 
(tunneling, say) between different localized systems, resulting in delocalized eigenstates
of $H_{eff}.$ From the numerical investigations discussed above, this appears to be the case in general,
but it is not obvious to what extent this argument should work, since in case of simple 
non-integrable models like kick-rotor (see Sec. 2), periodic drive actually induces localization 
by producing random destructive quantum interference (hence killing matrix elements), and the problem can 
be exactly mapped to Anderson localization. Other interesting scenarios are also being suggested and investigated
recently, see. e.g.,~\cite{2band,Saito}\\

\noindent
{\bf Stability of MBL towards High-frequency Drive:} This can be understood as follows. For fast drive
one can do Magnus expansion~\cite{Magnus} for $H_{eff}$ as follows.
\begin{eqnarray}
	H_{eff} &=& \sum_{n=0}^{\infty} H^{(n)}_{eff}, ~ {\rm where} \nonumber \\
	H^{(0)} &=& \frac{1}{T}\int_{0}^{T} H(t) dt \nonumber \\
	H^{(1)} &=& \frac{1}{2\!T(i\hbar)}\int_{\epsilon}^{\epsilon+T}dt_{1} 
	\int_{\epsilon}^{t_{1}}dt_{2}[H(t_{1}),H(t_{2})]\nonumber \\
	H^{(2)} &=& \frac{1}{3\!T(i\hbar)^{2}}\int_{\epsilon}^{\epsilon+T}dt_{1} 
	\int_{\epsilon}^{t_{1}}dt_{3}\left([H(t_{1}),[H(t_{2}),H(t_{3})]]\right. \nonumber \\
	&+& \left.[H(t_{3}),[H(t_{2}),H(t_{1})]]\right) \dots \\
\end{eqnarray}
For fast enough drive (small enough $T$), the series might converge, and 
one can keep only the leading order term $H^{(0)} = \frac{1}{T}\int_{0}^{T} H(t) dt$~\cite{Alessio_Rigol_PRX}. 
In that case of course $H_{eff}$ is trivially an MBL Hamiltonian, particularly when $\int_{0}^{T} H_{D}(t) dt = 0$ 
(which is the case considered in the above studies).

\subsection{In Presence of Mobility Edge}

\newcommand{\sz}[1]{\sigma_{#1}^{z}}
\newcommand{\sx}[1]{\sigma_{#1}^{x}}

\begin{figure}
	\begin{center}
	\includegraphics[width=0.4\textwidth]{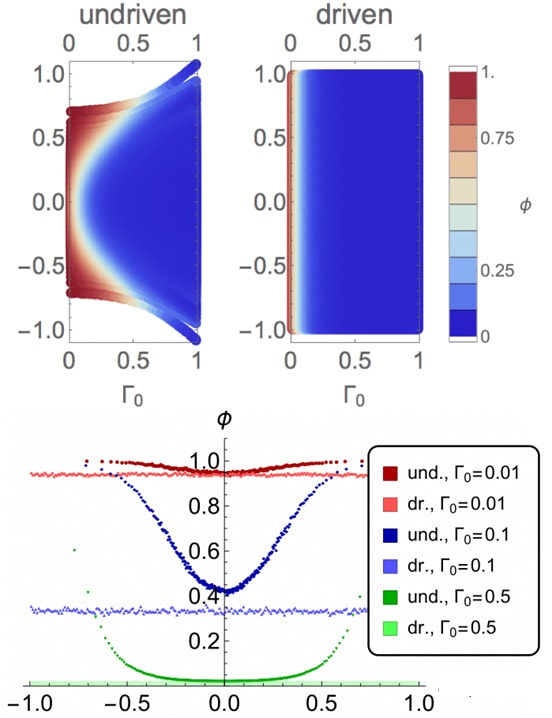}

	\protect\caption{Driving the QREM. The top left figure shows the participation ratio
		$\phi$ for the eigenstates of the undriven model, showing (energy/quasienergy on the y axis) 
		a mobile region (blue) surrounded by a localized region (red). Driving with
	frequency $\omega/J=0.1$ and amplitude $\delta/J=0.2$ (top right)
	causes all states at a given $\Gamma_{0}$ to become as delocalized
	as the least localized state at that $\Gamma_{0}$ in the undriven
	model. This is also shown in the bottom panel which shows $\phi$
	for $\Gamma_{0}=0.01,0.1,0.5$ (red, blue and green line, from top
	to bottom) in the absence (presence) of driving with darker (lighter)
	colour. The driven points always lie below the undriven points for
	the corresponding $\Gamma_{0}$. This is due to the strong mixing
	of all undriven eigenstates by the driving. All data in this figure
	is for $8$ spins and averaged over 1000 disorder realisations.
	(Fig. taken from~\cite{AL_AD_RM_PRL_2015}).
\label{fig:QREM}}
	\end{center}
\end{figure}
\begin{figure}[h!]
  \centering
	\includegraphics[scale=0.65]{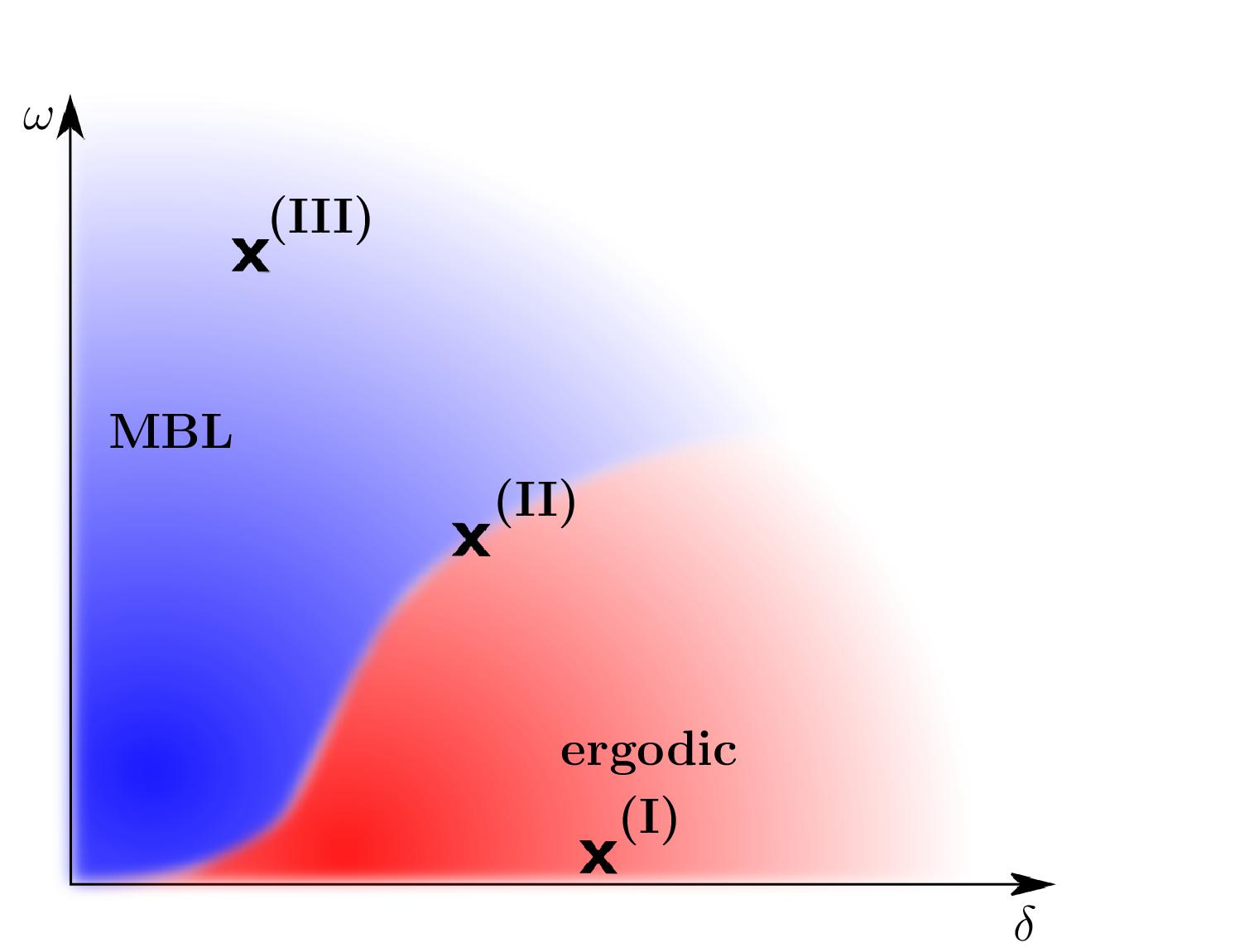}
	\caption{Proposed phase diagram for
the long time state of a driven strongly, disordered system as a function of
the driving frequency $\omega$ and strength $\delta$. Red (I)
and blue (III) indicate Floquet-ETH and
Floquet-MBL behavior, where the system approaches a fully-mixed,
``infinite-temperature'' state or remains localized, respectively.
At (II), heating leads to energy growing
logarithmically slowly with time (Fig.~\ref{fig:low-highfreq}(II)) over a broad time window.
(Fig. taken from~\cite{Jorge_MBL}).	}
	\label{fig:diagram}
\end{figure}
\begin{figure*}[h!]
	\begin{center}
	\includegraphics[scale=0.2]{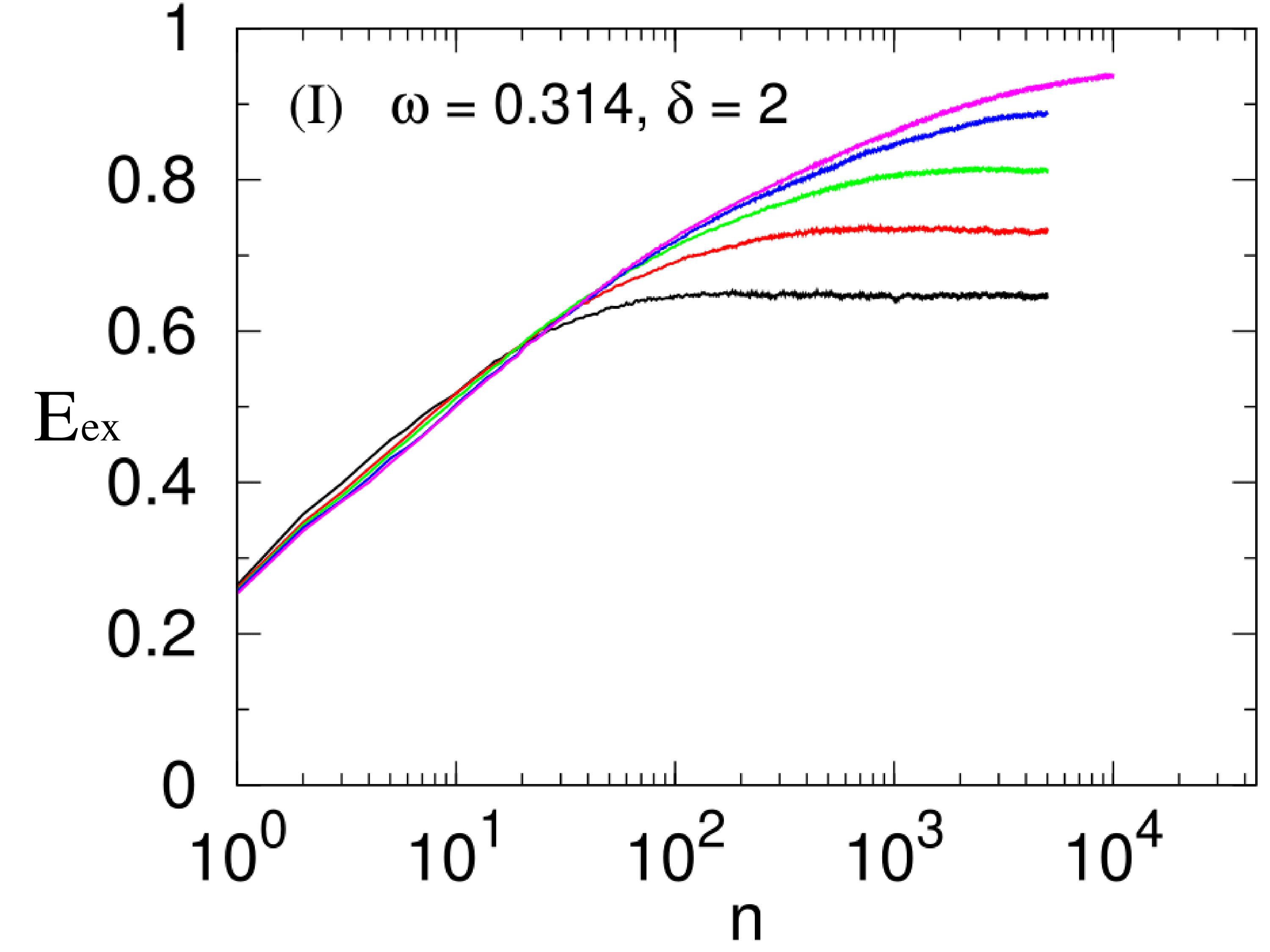}
	\includegraphics[scale=0.2]{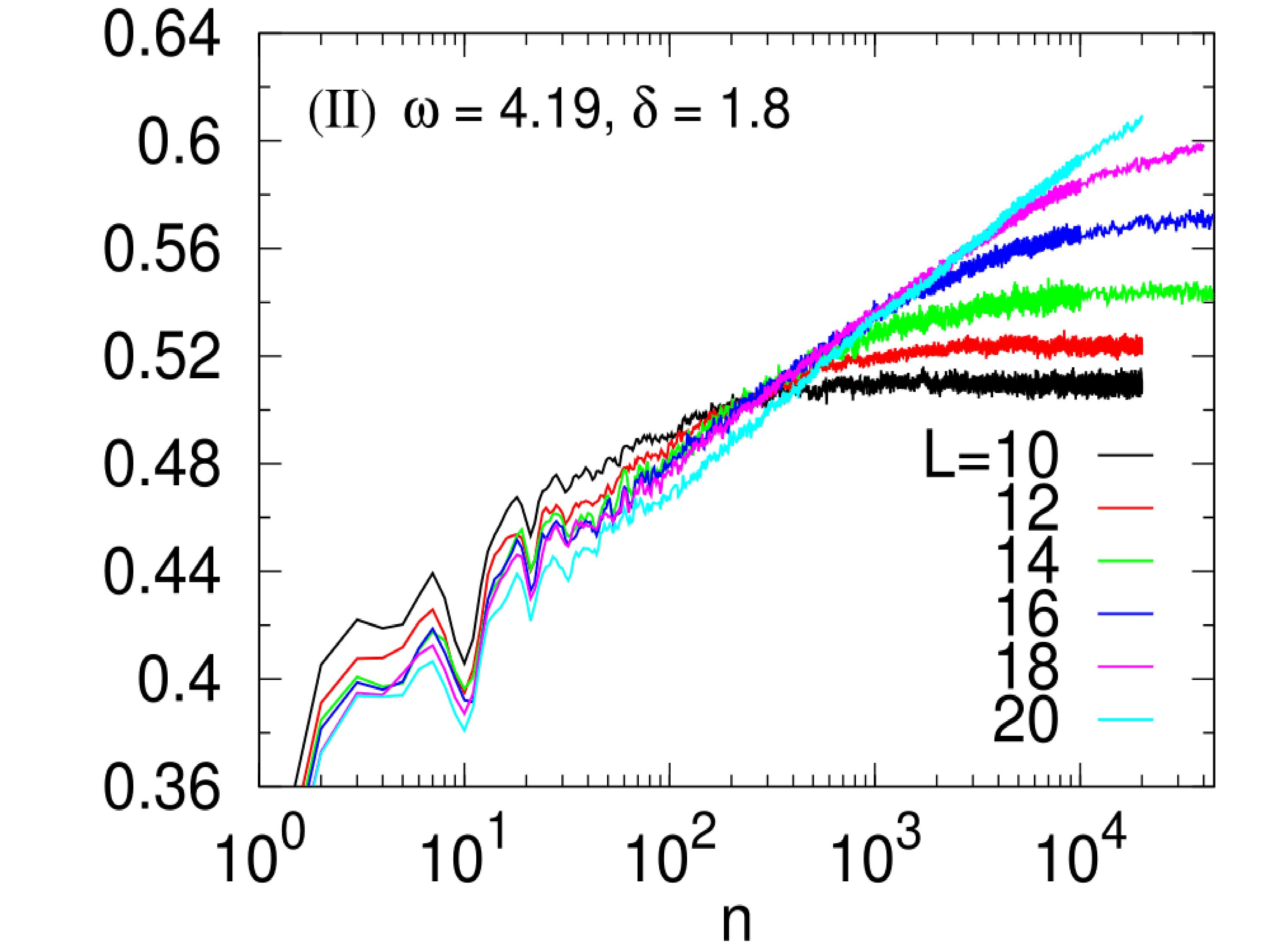}
	\includegraphics[scale=0.2]{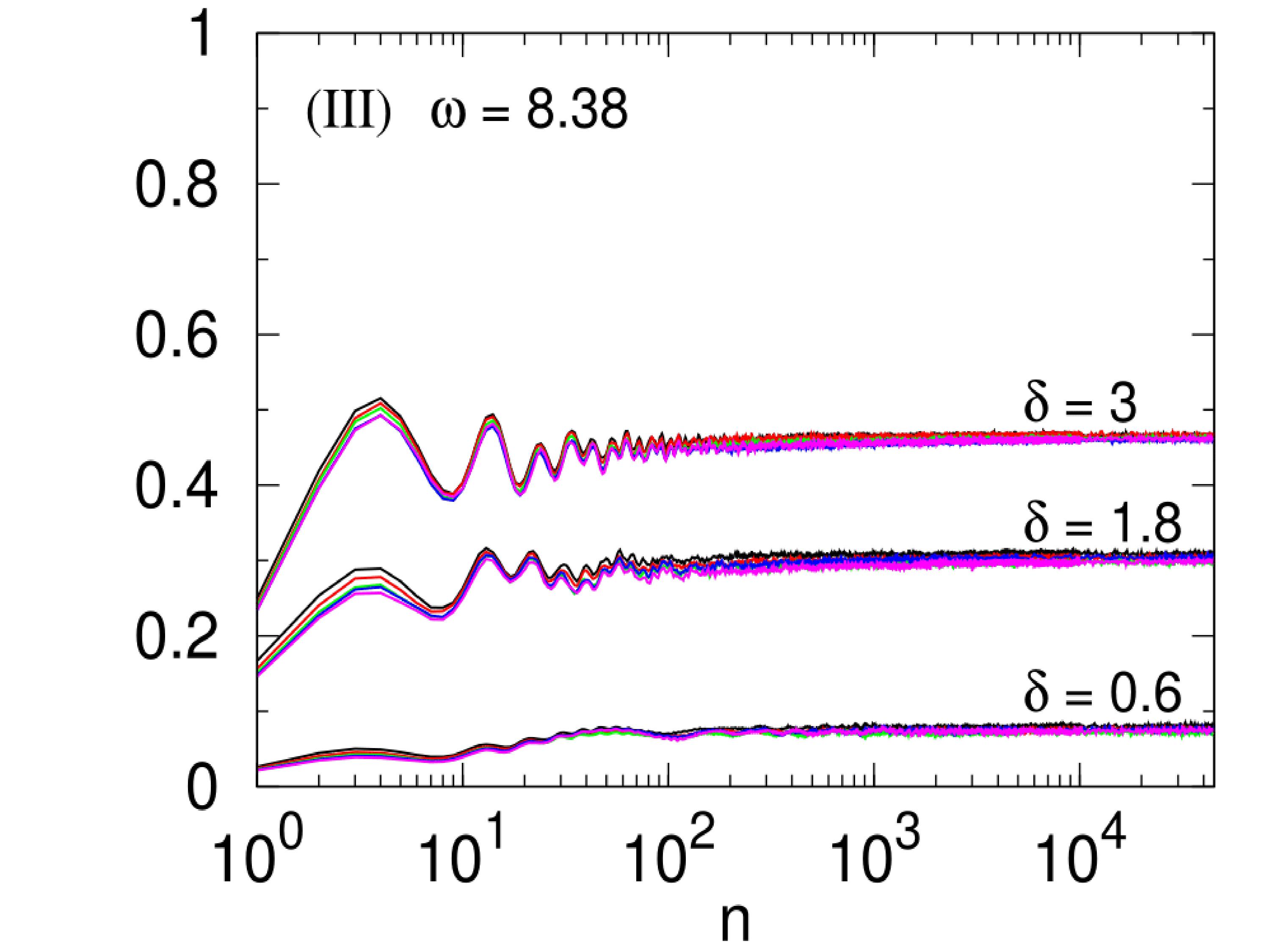}
	\caption{	
		Stroboscopic excess energy $E_{ex}(nT)$ in a strongly disordered system 
		with $\eta=5$ and $J_z=0.5$, corresponding to 
	three different regimes. In (I), an initially MBL system
delocalizes and heats up to a fully-mixed state.
	In the intermediate regime, (II), the system
heats up to the fully-mixed state, but logarithmically slowly. This
slow growth persists for longer times as we increase $L$.
	For Floquet-MBL, (III), driving does not
delocalize the system, leading instead to a localized long-time state which
has partially heated up to some intermediate energy. (Fig. taken from ~\cite{Jorge_MBL}).
	}
	\label{fig:low-highfreq}
	\end{center}
\end{figure*}

{\bf The QREM as a case study:} We now turn to the case in which a 
mobility edge is present in the undriven spectrum. 
The central result is based on the observation~\cite{AL_AD_RM_PRE_2014} 
that a periodic perturbation acting on a system couples each undriven state to states 
spread uniformly throughout the spectrum of $H_{0}$. As a result, if part of the 
spectrum corresponds to delocalized eigenstates then all eigenstates of $H_{eff}$ 
will necessarily be delocalized. Whether there exists mobility edge in 
local models of MBL is debatable~\cite{DeRoeck}, but delocalization for all values of $\omega$ 
has been numerically confirmed by studying the Quantum Random Energy Model 
(QREM), recently studied in Ref.~\cite{Laumann:2014vd} 
where it was shown to have a mobility edge. The model is defined for $N$ Ising spins with the Hamiltonian 
$H = E\left(\left\{\sz{j} \right\}\right) - \Gamma \sum_j \sx{j}$, where $E$ is a 
random operator diagonal in the $\sigma^z$ basis (that is, it assigns a random 
energy to each spin configuration) and $\Gamma$ a transverse field. Extensivity 
of the many-body spectrum is satisfied if the random energies are drawn from a 
distribution $P\left(E\right)=\frac{1}{\sqrt{\pi N}}\exp\left(-E^{2}/N\right)$.

The diagnostic of localization used here is the participation ratio (PR), defined for the state 
$|\psi\rangle$ with respect to the Fock basis $\{|n\rangle\}$ 
as $\phi=\sum_{n}\mbox{\ensuremath{\left|\left\langle n\right|\left.\psi\right\rangle \right|}}^{4}$ 
with $n$ enumerating Fock states. $\phi$ approaches unity for a state localized on a single Fock state and $2^{-N}$ 
for one fully delocalized in Fock space. The leftmost panel in Fig.~\ref{fig:QREM} shows the average 
$\phi$ versus energy (scaled with system size) of the 256 eigenstates of an undriven $N=8$ system averaged 
over 1000 disorder realisations, demonstrating the existence of a mobility edge.

The system is also driven by modulating 
$\Gamma\left(t\right)=\Gamma_{0}\left(1+\delta\tilde{\delta}\left(t\right)\right)$, 
$\tilde{\delta}\left(t\right)=+1(-1)$ for the first (second) half of the period with an amplitude 
$\delta=0.2$ and frequency $\omega=2\pi/T=0.1$. The PR of the eigenstates of $H_{eff}$ 
are shown in the second panel of  Fig.~\ref{fig:QREM}. As expected, periodic driving causes delocalization of 
the entire spectrum so long as part of the undriven spectrum at the same $\Gamma_{0}$ is delocalized.

\subsection{Energy Absorption:} 
Energy absorption by an MBL system under periodic drive has been studied extensively in~\cite{Jorge_MBL}.
The model studied was spin-1/2 XXZ chain in a disordered longitudinal field under monochromatic drive with
period $T = 2\pi/\omega:$
\begin{eqnarray}
H(t) & = & H_{0} + H_D(t)   \label{eq:ham_drive} \\
	H_{0} &=& J_{\perp}\sum_{i=0}^L (S^x_iS^x_{i+1} + S^y_iS^y_{i+1}) + J_z\sum_{i=0}^L S^z_iS^z_{i+1}\\
	& & +  \sum_{i=0}^L h^z_iS^z_i,
\end{eqnarray}
where $h^z_i\in[-\kappa,\kappa]$, $J_{\perp},J_z\ge0$,
and with driving
\begin{align}
H_{D}(t)= -\delta\cos{\omega t}\sum_{i=0}^L (-1)^i S^z_i .
  \label{eq:ham_dri}
\end{align}

The static part $H_0$ is known to be MBL  for $J_z\neq0$
and sufficiently strong disorder $\kappa>\kappa_c.$ Starting from the
ground state of $H_{0}$ at $t=0,$ real-time
dynamics of ``rescaled excess energy density" defined as
\begin{align}
	E_{ex}(nT) = \frac{\langle\psi| H(nT)|\psi\rangle-E_{\mathrm{min}}}{\overline{E}-E_{\mathrm{min}}},
\end{align}
with $\overline{E}=D_H^{-1}\mathrm{tr}[H(0)]$ and $D_H$ being the
Hilbert space dimension, so that $E_{ex}=0$ in the ground state of $H(0)$, 
while $E_{ex}=1$ for a state uniformly delocalized among all eigenstates of $H_{0}.$ This is 
the ``infinite temperature" like scenario observed when a disorder free interacting system is
driven periodically~\cite{AL_AD_RM_PRE_2014}. 

For strong disorder ($\kappa > \kappa_{c}$),  
a qualitative phase-diagram depending on $\omega$ and $\delta$ (Fig.~\ref{fig:diagram}) has been
proposed. Three regimes has been identified (marked in the figure) as follows. There are two
regimes (I and III), which are known from earlier works~\cite{AL_AD_RM_PRL_2015,Abanin_MBL} (see
also Sec.~\ref{No_Mobility_Edge}).
In regime (I) an initially MBL state delocalizes and heats up to a state in which the system locally looks as if it is at infinite temperature 
(i.e., the expectation value of the local observables over the state equals to that over an infinite temperature ensemble). 
In (III), the drive fails to delocalize the system, but only pumps some energy into it. The energy 
of the system thus settles to some intermediate average value. However, the numerics seems to 
suggest there is another intermediate regime (II), where the system tend to reach the infinite
temperature like scenario, but only logerithmically slowly. Sample of real time behaviour of these
regimes are given in Fig.~\ref{fig:low-highfreq}. Further technical details, particularly
those relevant for validation of the numerical results are given in detail in~\cite{Jorge_MBL}.

\section{Summary and Outlook} In this article we illustrate that periodic drive can induce both freezing/localization and 
unfreezing/delocalization dynamically when applied to closed quantum systems with many-degrees of freedom depending on the
circumstances. It has been argued that quantum interference plays an important role in all these phenomena,
though the intuitive pictures consistent with different circumstances are quite varied. For example, 
while in the case of DMF the simplest physical picture of freezing
seems to consist of renormalization of the Hamiltonian by the periodic drive resulting in suppression of dynamics,
dynamical localization in a periodically kicked rotor 
can be understood more easily by mapping it to the Anderson
localization problem, implying that the drive induced dynamical randomness 
has similar localizing effect as quenched disorder in static problem. 
Yet, in an inherently localized many-body system with interaction, periodic drive at low frequencies can cause {\it delocalization}. 
There the picture is, the drive generates matrix elements connecting the spatially 
isolated localized parts of the undriven MBL system. The system then  eventually heats up till it 
reaches a state which looks like an infinite temperature state when expectation values of local observables 
are measured. Interestingly, it seems no noticeable dynamical localization is observed in periodically driven MBL
in the low $\omega$ regime (at least, not sufficiently strong to exhibit any noticeable freezing effect), unlike
that observed in single-body non-integrable quantum chaotic system like the kick-rotator.
The issue of energy absorption under external drive in system not attached to a bath is
still a broad open question, and are being pursued under different 
drive protocols~\cite{R2a,R2b}.

A number of interesting questions present themselves. \\

\noindent
For example, whether under high frequency drive, mechanism of dynamical localization steps and lends an 
MBL phase greater stability? \\

\noindent
What happens if one drives an integrable models (which can be mapped to non-interacting fermions) 
with very low frequencies, and the Magnus expansion breaks down? Moreover
Does one achieve effective infinite temperature thermalization
scenario there? If yes, then how does the transition/crossover from PGE to thermal phase takes place ? If not, then certainly
interaction kills it (evidences indicate that MBL gives way to ergodic phases when driven with low enough 
frequencies). Then the behavior of MBL-ergodic crossover as a function of interaction strength would be interesting. \\

\noindent
Can one expect Periodic (generalized) Gibbs' Ensemble as a local description of the asymptotic states in periodically driven integrable systems which cannot be mapped to free fermions? 
This seems plausible, but to our knowledge there is no general proof of existence of extensive number of stroboscopic conserved quantities necessary for this. 
The same (open) question appears interesting in the context of DMF, in particular, whether extreme freezing points can occur in such integrable systems.\\

\noindent
One might wonder what happens when a bath is weakly coupled to a periodically driven system. 
The effect of quantum interference would be affected by external decoherence and dissipation, 
and of course the asymptotic behaviour cannot be expected to be determined by a statistical 
average of the property of $H_{eff}$ (system would not go to a diagonal ensemble in the 
eigenbasis of $H_{eff}$ in general). The intuitive pictures developed here might have to be 
revised significantly, and emergence of fundamentally new pictures are not unlikely.
\\

\noindent
{\bf Acknowledgements:} AD acknowledges collaborations with 
S. Bhattacharyya, S. Dasgupta, A. Lazarides, R. Moessner and A. Roy in 
various works covered in this article. AH and AD thankfully acknowledge support from DST-MPI
partner group program ``Spin liquids: correlations, dynamics
and disorder" between MPI-PKS (Dresden) and IACS (Kolkata), and the Visitor's Program of 
MPI-PKS for a visit to MPI-PKS, during which many interesting discussions on the subject took place.

\end{document}